\documentclass[hyper]{JHEP} % 10pt is ignored!
\usepackage{epsfig}
\newcommand\fverb{\setbox\pippobox=\hbox\bgroup\verb}
\newcommand\fverbdo{\egroup\medskip\noindent%
            \fbox{\unhbox\pippobox}\ }

\newcommand\fverbit{\egroup\item[\fbox{\unhbox\pippobox}]}
\newbox\pippobox
\title{Three Spin Spiky Strings in $\beta$-deformed Background}
\author{Kamal L. Panigrahi, Pabitra M. Pradhan and Pratap K. Swain \\
Department of Phyisics and Meteorology, \\
Indian Institute of Technology Kharagpur,\\
Kharagpur-721 302, INDIA \\
E-mail: \email{panigrahi,ppabitra,pratap@phy.iitkgp.ernet.in}}
\abstract{We study rigidly rotating strings in $\beta$-deformed
$AdS_5\times S^5$ background with one spin along AdS$_5$ and
two angular momenta along $S^5$. We find the spiky string
solutions and present the dispersion relation among various
charges in this background. We further generalize the result to
the case of four angular momenta along $AdS_5 \times
S_{\gamma}^5$.} 

\begin{document}
\section{Introduction}
AdS/CFT correspondence \cite{Maldacena:1997re} relates the
spectrum of free strings on AdS$_5\times$ S$^5$ with the spectrum
of operator dimensions in the $N = 4$ Supersymmetric Yang-Mills
(SYM) theory in four dimensions. This mapping is highly nontrivial
and challenging because of our lack of understanding of the full
string theory spectrum. Hence, it is instructive to look at both
the gauge and gravity theories at certain limits such as large
angular momentum limit and then compare the spectrum on both
sides. Further, $N = 4$ SYM theory can be described by integrable
spin chain model where the anomalous dimension of the gauge
invariant operators were found \cite{Minahan:2002ve}. It was
further noticed that string theory has as integrable structure in
the semiclassical limit and the anomalous dimension in the N = 4
SYM can be derived from the relation between conserved charges of
the worldsheet solitonic string solution of the dual string theory
on AdS$_5\times$ S$^5$ background.

In the past few years the integrability of both string theory and
gauge theory side has played a key role in proving the duality
conjecture better. The most frequently studied cases were rotating
and pulsating strings solutions in certain limits of spin waves in
long-wave approximation and interesting observations were made in
\cite{Tseytlin:2003ii}, \cite{Dimov:2004xi},
 \cite{Smedback:1998yn}, \cite{Beisert:2004ry}, \cite{Tseytlin:2004xa},
\cite{Plefka:2005bk}. Another interesting case is the low lying
spin states which are equivalent to the so called magnon states.
In this connection, Hofman and Maldacena in \cite{Hofman:2006xt}
have derived a mapping between particular state (magnon) of spin
chain with the semiclassical string states on $R_t \times S^2$ .
Further it was realized that magnons are special cases of more
general solutions known as spikes and the dual gauge theory
operators have been analyzed in detail in
\cite{Kruczenski:2004wg}. It was also observed in
\cite{Ishizeki:2007we} that both giant magnon and single spike
solutions can be viewed as two different limits of the same
rigidly rotating strings on S$^2$ and S$^3$. In this connection, a
large class of multispin spiky string and giant magnon solutions
have been studied, in various backgrounds including the orbifolded
and non-AdS backgrounds, for example in \cite{spike}

%\cite{Bobev:2006fg}, \cite{Kluson:2007qu}, \cite{Lee:2008sk},
%cite{Lee:2008ui}, \cite{Ryang:2008rc}, \cite{Jain:2008mt},
%\cite{Benvenuti:2008bd}, \cite{Biswas:2011wu}.

We are interested in studying a class of spiky string solution in
the so called Lunin-Maldacena background \cite{Lunin:2005jy}. The
integrability of Lunin-Maldacena background has been studied in
\cite{Roiban:2003dw}, \cite{Berenstein:2004ys},
\cite{Frolov:2005ty}. The giant magnon and single spike solutions
are studied in detail including the integrable models, for example
in \cite{Chu:2006ae}, \cite{Bobev:2007bm}.
%Another class of magnon solution found in
%\cite{Minahan:2006bd}, \cite{Ryang:2006yq} with two and three spin
%respectively having spin both in AdS and spherical part of $AdS_5
%\times S^5$ background. %The giant magnon relation found in
%\cite{Hofman:2006xt} is:
%\begin{equation}
%E-J = \frac{\sqrt{\lambda}}{\pi} \sin{\frac{p}{2}},
%\end{equation}
%where $p$ is the magnon momentum. In string theory side $p$
%is defined as angular difference of the magnon. This magnon relation
%is extended in case of multispin states both on the string and spin chain side.
%In case of two spin the magnon relation is:
%\begin{equation}
%E - J_1 = \sqrt{J_2^2 + \frac{\lambda}{\pi^2}\sin^2{\frac{p}{2}}}.
%\end{equation}

More recently in \cite{Giardino:2011dz}, \cite{Panigrahi:2011be},
\cite{Giardino:2011uc}, more general class of solutions with three
divergent angular momenta have been studied and interesting
dispersion relations among various conserved charges have been
obtained. In the present paper, we wish to generalize the result
of \cite{Ryang:2006yq} in a $\beta$-deformed background with
one(two) spin in AdS and two spins on the  deformed sphere.
Knowing such solutions on the gravity  side will definitely help
us in finding out the nature of corresponding operators on the
gauge theory side.

The rest of the paper is organized as follows. In section-2 we
write the relevant part of the Lunin-Maldacena background which
will be useful for studying the rigidly rotating string on this
background. In section 3, we study the rotating open string in
$AdS_3 \times S_{\gamma}^3$ backgrounds with one spin along the
AdS$_3$ and two angular momenta along the deformed sphere. We
compute all the conserved charges and find two limiting cases
corresponding to giant magnon and single spike solutions. We write
down the dispersion relation among various divergent momenta in
both cases. We further generalize the above solutions to the case of 
rotating string with two spins along the AdS$_3$ and two angular
momenta along the deformed sphere. We write down the corresponding
dispersion relation for the giant magnon solution. Finally in
section 4, we conclude with some remarks.

\section{$\beta$-deformed $AdS_5 \times S^5$ background}
Here we present the general background for $\beta$-deformed $AdS_5
\times S^5$ found by Lunin and Maldacena\cite{Lunin:2005jy}. This
background is obtained from pure $AdS_5 \times S^5$ by a series of
STsTS transformations \cite{Frolov:2005dj}, which is dual to the
Leigh-Strassler marginal deformations of $N = 4$ SYM
\cite{Leigh:1995ep}. The deformed parameter $\beta = \gamma +
i\sigma_d$ in general is a complex number, but here we restrict
$\beta$ to its real part only. Thus, the relevant metric component
of the supergravity background dual to real $\beta$-deformations
of $N = 4$ SYM is:
\begin{equation}
ds^2 = R^2\left(ds^2_{AdS_5} + \sum_{i=1}^3(
d{\mu_i}^2 + G{\mu_i}^2d\phi_i^2) + \tilde{\gamma}^2G
\mu_1^2\mu_2^2\mu_3^2(\sum_{i=1}^3d\phi_i^2\right),
\nonumber \\ \label{1}
\end{equation}
which also have the dilaton, Ramond-Ramond (RR) and
Neveu-Schwarz-Neveu-Schwarz (NS-NS) fields.

The antisymmetric form of the $B$-field relevant for our classical
string analysis is:
\begin{equation}
B = R^2\tilde{\gamma} G(\mu_1^2\mu_2^2d\phi_1d\phi_2 +
\mu_2^2\mu_3^2d\phi_2d\phi_3 + \mu_1^2\mu_3^2d\phi_1d\phi_3),
\nonumber \\ \label{2}
\end{equation}
where
\begin{eqnarray}
\tilde{\gamma} &=& R^2\gamma,~~~ R^2 = \sqrt{4\pi g_s N}, \cr &
\cr G &=& \frac{1}{1 + \tilde{\gamma}^2(\mu_1^2\mu_2^2 +
\mu_2^2\mu_3^2 + \mu_1^2\mu_3^2)}, \cr & \cr \mu_1 &=&
\sin\theta\cos\psi, ~~~ \mu_2 = \cos\theta,~~~ \mu_3 =
\sin\theta\sin\psi. \nonumber \\ \label{3}
\end{eqnarray}
\section{Semiclassical Strings on AdS$_3 \times$ S$^3_{\gamma}$}

We restrict the motion of the string on  $AdS_3 \times
S_{\gamma}^3$ $\subset$ $AdS_5 \times S_{\gamma}^5$. This space
can be achieved by putting $\mu_3 =0$, $\phi_3 = 0$ i.e, $\psi =
0$, $\phi_3 = 0$ in (\ref{1}) - (\ref{3}). Thus the metric
components of the deformed $AdS_3 \times S_{\gamma}^3$ background
is:
\begin{equation}
ds^2 = -\cosh^2\rho dt^2 + d\rho^2 + \sinh^2\rho d\phi^2 +
d\theta^2 + G\sin^2\theta d\phi_1^2 + G\cos^2\theta d\phi_2^2,
\nonumber \\ \label{4}
\end{equation}
where $G = \frac{1}{1 + \tilde{\gamma}^2\sin^2\theta\cos^2\theta}$
and the relevant non-zero component of B-field due to the series of T-duality
transformations is given by
\begin{equation}
B_{\phi_1\phi_2} = \tilde{\gamma} G \sin^2\theta \cos^2\theta.
\nonumber \\ \label{5}
\end{equation}
The Polyakov action for the fundamental string in this background is
given by
\begin{eqnarray}
S &=& \frac{T}{2}\int d\tau d\sigma\Big[-\cosh^2\rho
(\dot{t}^2 - {t^{\prime}}^2) + \dot{\rho}^2 - {\rho^{\prime}}^2 +
\sinh^2\rho(\dot{\phi}^2 - {\phi^{\prime}}^2) +
\dot{\theta}^2 - {\theta^{\prime}}^2 +
\cr & \cr
&&G\sin^2\theta(\dot{\phi_1}^2 - {\phi_1^{\prime}}^2)
+ G\cos^2\theta(\dot{\phi_2}^2 - {\phi_2^{\prime}}^2)  +
2\tilde{\gamma} G \sin^2\theta\cos^2\theta(\dot{\phi_1}\phi_2^{\prime} -
\phi_1^{\prime}\dot{\phi_2})\Big],
\nonumber \\ \label{6}
\end{eqnarray}
where the `dot' and `prime' denote the derivatives with respect to
$\tau$ and $\sigma$ respectively and $T =
\frac{\sqrt{\lambda}}{2\pi}$, where $\lambda$ is the `t Hooft coupling constant.
We take the following anstaz for the rotating open string
\begin{eqnarray}
t &=& \tau + F_1(y), ~~~ \rho = \rho(y), ~~~ \phi =
\omega_1(\tau + F_2(y)),
\cr & \cr
\phi_1 &=& \tau + F_3(y), ~~~ \theta = \theta(y), ~~~
\phi_2 = \omega_2(\tau + F_4(y)),
\nonumber \\ \label{7}
\end{eqnarray}
where $y = a\sigma - b\tau$.
Solving the equations of motion for $t, \phi, \phi_1$ and $\phi_2$, we
have the following expression for $F_1, F_2, F_3$ and $F_4$
\begin{eqnarray}
F_{1y} &=& \frac{1}{a^2 - b^2}\left(\frac{A_1}{\cosh^2\rho} - b\right),
\cr & \cr
F_{2y} &=& \frac{1}{a^2 - b^2}\left(\frac{A_2}{\sinh^2\rho} - b\right),
\cr & \cr
F_{3y} &=& \frac{1}{a^2 - b^2}\left(\frac{A_3}{G\sin^2\theta} - b -
a\tilde{\gamma}\omega_2\cos^2\theta\right),
\cr & \cr
F_{4y} &=& \frac{1}{a^2 - b^2}\left(\frac{A_4}{G\cos^2\theta} - b +
\frac{a\tilde{\gamma}}{\omega_2}\sin^2\theta\right),
\nonumber \\ \label{8}
\end{eqnarray}
where $F_y = \frac{\partial F}{\partial y}$.

Now, the two virasoro constraints $T_{\tau\sigma} = 0$
and $T_{\tau\tau} + T_{\sigma\sigma}  = 0$ give the following two equations
\begin{eqnarray}
\rho_y^2 + \theta_y^2 &=& \cosh^2\rho(-\frac{1}{b}F_{1y} +
F_{1y}^2) - \omega_1^2 \sinh^2\rho(-\frac{1}{b}F_{2y} +
F_{2y}^2)
\cr & \cr
&&- G\sin^2\theta(-\frac{1}{b}F_{3y} + F_{3y}^2) -
\omega_2^2G\cos^2\theta(-\frac{1}{b}F_{4y} + F_{4y}^2)
\cr & \cr
\rho_y^2 + \theta_y^2 &=& \cosh^2\rho(\frac{1}{a^2 + b^2}
-\frac{2b}{a^2 + b^2}F_{1y} + F_{1y}^2)
- \omega_1^2 \sinh^2\rho(\frac{1}{a^2 + b^2}
-\frac{2b}{a^2 + b^2}F_{2y} + F_{2y}^2)
\cr & \cr
&&- G\sin^2\theta(\frac{1}{a^2 + b^2}
-\frac{2b}{a^2 + b^2}F_{3y} + F_{3y}^2)
- \omega_2^2G\cos^2\theta(\frac{1}{a^2 + b^2}
-\frac{2b}{a^2 + b^2}F_{4y} + F_{4y}^2),\nonumber \\
\label{9}
\end{eqnarray}
Using (\ref{8}) in (\ref{9}) we get the following relation among various constants
\begin{equation}
-A_1 + \omega_1^2 A_2 + A_3 + \omega_2^2 A_4 = 0
\nonumber \\ \label{10}
\end{equation}
Now, the equation of motion for $\rho$ becomes
\begin{equation}
(a^2-b^2)\rho_{yy} + \frac{1}{a^2-b^2} \sinh\rho\cosh\rho
\left(\frac{A_1^2}{\cosh^4\rho} - a^2 - \frac{\omega_1^2 A_2^2}
{\sinh^4\rho} + \omega_1^2 a^2\right) = 0 \ ,
\nonumber \\ \label{11}
\end{equation}
where $\rho_{yy} = \frac{\partial^2 \rho}{\partial y^2}$.
We can compute equation of motion for $\theta$ from the first
Virasoro constraint (\ref{9}) which is given by the relation
\begin{eqnarray}
\theta_y^2 &=& -\rho_y^2 + \cosh^2\rho(-\frac{1}{b}F_{1y} +
F_{1y}^2) - \omega_1^2 \sinh^2\rho(-\frac{1}{b}F_{2y} +
F_{2y}^2)
\cr & \cr
&&- G\sin^2\theta(-\frac{1}{b}F_{3y} + F_{3y}^2) -
\omega_2^2G\cos^2\theta(-\frac{1}{b}F_{4y} + F_{4y}^2) \ .
\nonumber \\ \label{t}
\end{eqnarray}
Once we get the form of $\rho_y$ then we will be able to compute
$\theta_y$. The conserved charges are
\begin{eqnarray}
E &=& T~\int d\sigma~\cosh^2\rho(1 - bF_{1y}), \cr & \cr S &=&
\omega_1T~\int d\sigma~ \sinh^2\rho (1 - bF_{2y}), \cr & \cr J_1
&=& T~\int d\sigma~G\sin^2\theta (1 - bF_{3y}), \cr & \cr J_2 &=&
\omega_2T ~\int d\sigma~G\sin^2\theta (1 - bF_{4y}). \nonumber \\
\label{cc}
\end{eqnarray}
\subsection{Giant Magnon Solution}
For finding out the giant magnon solution, we choose the
integation constants as  $A_1 = b, A_2 = 0, A_3 =b$ and $A_4 =0$.
The solution of equation (\ref{11}) becomes
\begin{equation}
\rho_y^2 = \frac{1}{(a^2-b^2)^2}
\left(a^2(1-\omega_1^2) -\frac{b^2}{\cosh^2\rho}\right)\sinh^2\rho.
\nonumber \\ \label{12}
\end{equation}
Using (\ref{12}) and the above integration constants in (\ref{t}), we get the
following expression for $\theta_y$
\begin{equation}
\theta_y^2 = \frac{1}{(a^2-b^2)^2}\left[a^2(1-\omega_2^2)
\cos^2\theta +b^2 - \frac{b^2}{G\sin^2\theta} + 2ab\tilde{\gamma}\omega_2
\cos^2\theta\right] \ .
\nonumber \\ \label{13}
\end{equation}
Note that the above equation can be rewritten as
\begin{equation}
\theta_y = \frac{\Omega_0}{a^2-b^2}\cot\theta
\sqrt{\sin^2\theta - \sin^2\theta_0} \ ,
\nonumber \\ \label{14}
\end{equation}
where, $\sin\theta_0 = \frac{b}{\Omega_0}$, and
$\Omega_0 = \sqrt{a^2 - (a\omega_2 -b\tilde{\gamma})^2}$.
Now the conserved charges (\ref{cc}) become
\begin{eqnarray}
E &=& \frac{T}{a^2-b^2}\int d\sigma~(a^2\cosh^2\rho - b^2), \cr &
\cr \frac{S}{\omega_1} &=& \frac{T}{a^2-b^2}\int d\sigma~
a^2\sinh^2\rho, \cr & \cr J_1 &=& \frac{T}{a^2-b^2}\int
d\sigma~(a^2\sin^2\theta - b^2), \cr & \cr \frac{J_2}{\omega_2}
&=& \frac{T}{a^2-b^2} \int d\sigma~ a^2(1 -
\frac{b\tilde{\gamma}}{a\omega_2})\cos^2\theta. \nonumber \\
\label{15}
\end{eqnarray}
It is clear from the above expressions that we have the following
relation among various conserved charges
\begin{equation}
E - J_1 = \frac{S}{\omega_1} + \frac{J_2^{\prime}}{\omega_2} \ ,
\nonumber \\ \label{16}
\end{equation}
where $J_2^{\prime} =
\frac{a\omega_2}{\sqrt{a^2-{\Omega_0}^2}}J_2$. As the conserved
charges are divergent, we use the same regularization technique as
in \cite{Ryang:2006yq} to remove the divergent part of the
conserved charges. Let us write
\begin{eqnarray}
\frac{S}{\omega_1} = \frac{2Ta}{a^2-b^2}\int_{\infty}^0~
d\rho \frac{\sinh^2\rho}{\rho_y}
= \frac{2T}{\sqrt{1-\omega_1^2}} \int_{1}^{\infty}~
dz \frac{z}{\sqrt{z^2 - z_0^2}} \ ,
\nonumber \\ \label{17}
\end{eqnarray}
where $z = \cosh\rho$ and $z_0 = \cosh\rho_0 =
\frac{b}{a\sqrt{1-\omega_1^2}}$.
%\begin{equation}
%\frac{S}{\omega_1} = \frac{\sqrt{\lambda}}{\pi}
%\frac{1}{\sqrt{1-\omega_1^2}}
%left(\infty -{\sqrt{1-z_0^2}}\right)
%\nonumber \\ \label{18}
%\end{equation}
Subtracting the divergent part of the integral (\ref{17}), we have the regulated
value given as
\begin{equation}
\frac{S_{reg}}{\omega_1} = -\frac{\sqrt{\lambda}}{\pi}
\sqrt{\frac{1-z_0^2}{1-\omega_1^2}} \ .
\nonumber \\ \label{19}
\end{equation}
From the above expression, we find the following relation
\begin{equation}
\frac{S_{reg}}{\omega_1} = - \sqrt{S_{reg}^2 +
\frac{\lambda}{\pi^2}(1-z_0^2)} \ .
\nonumber \\ \label{20}
\end{equation}
Further, the time difference between two end points of the open string is given by
\begin{eqnarray}
\Delta t &=& -\frac{2b}{a\sqrt{1-\omega_1^2}}
\int_{-\infty}^{\infty}~d\rho\frac{\tanh\rho}{\sqrt{\cosh^2\rho - \frac{b^2}{a^2(1-\omega_1^2)}}}
\nonumber \\
&=& - 2 \tan^{-1}\frac{z_0}{\sqrt{1-z_0^2}} \ .
\nonumber \\ \label{21}
\end{eqnarray}
Now the equation (\ref{20}) can be written in the following form
\begin{equation}
\frac{S_{reg}}{\omega_1} = - \sqrt{S_{reg}^2 +
\frac{\lambda}{\pi^2}\cos^2\frac{\Delta t}{2}} \ .
\nonumber \\ \label{22}
\end{equation}
Further, the angle difference between two end points of the open string is given by
\begin{eqnarray}
\frac{\Delta\phi_1}{2}
= \int_{\theta_0}^{\frac{\pi}{2}}~d\theta
\frac{\tilde{\gamma}\sqrt{a^2-\Omega_0^2}}{\Omega_0}\cot\theta
\frac{\sin^2\theta + \frac{b}{\tilde{\gamma}\sqrt{a^2-\Omega_0^2}}}
{\sqrt{\sin^2\theta - \sin^2\theta_0}}
= \frac{\pi}{2} - \theta_0 +
\frac{\tilde{\gamma}\sqrt{a^2-\Omega_0^2}}{\Omega_0}
\cos\theta_0.
\nonumber \\ \label{23}
\end{eqnarray}
Now, we find the giant magnon dispersion relation as
\begin{eqnarray}
(E - J_1)_{reg} =  - \sqrt{S_{reg}^2 +
\frac{\lambda}{\pi^2}\cos^2\frac{\Delta t}{2}} +
\sqrt{J_2^2 +
\frac{\lambda}{\pi^2}\sin^2\frac{\Delta \phi_1}{2}}.
\nonumber \\ \label{24}
\end{eqnarray}
This expression matches with that of \cite{Ryang:2006yq} even if
we are dealing with $\beta$-deformed background and has implicit
dependence on the deformation parameter $\tilde{\gamma}$ in the
definition of $\Delta \phi_1$.
\subsection{Single spike Solution}

To obtain the single spike solution, we chose the integration
constants as: $A_1 = \frac{a^2}{b} = A_3$  and $A_2 = 0 = A_4$.
The solution of equation (\ref{11}) now becomes
\begin{equation}
\rho_y^2 = \frac{1}{(a^2-b^2)^2}
\left(a^2(1-\omega_1^2) -\frac{a^4}{b^2\cosh^2\rho}\right)\sinh^2\rho.
\nonumber \\ \label{25}
\end{equation}
Using (\ref{25}) and the above integration constants in (\ref{t}), we have the
follwing expression for $\theta_y$:
\begin{equation}
\theta_y = \frac{a\Omega_1}{a^2-b^2}\cot\theta
\sqrt{\sin^2\theta - \sin^2\theta_1},
\nonumber \\ \label{26}
\end{equation}
where, $\sin\theta_1 = \frac{a}{b\Omega_1}$, and
$\Omega_1 = \sqrt{1 - (\omega_2 -\frac{a\tilde{\gamma}}{b})^2}$.
Thus the conserved charges (\ref{cc})  becomes:
\begin{eqnarray}
E &=& \frac{T}{a^2-b^2}\int d\sigma~a^2\sinh^2\rho, \cr & \cr
\frac{S}{\omega_1} &=& \frac{T}{a^2-b^2}\int d\sigma~
a^2\sinh^2\rho, \cr & \cr J_1 &=& -\frac{T}{a^2-b^2}\int
d\sigma~a^2\cos^2\theta, \cr & \cr \frac{J_2}{\omega_2} &=&
\frac{T}{a^2-b^2} \int d\sigma~ a^2(1 -
\frac{a\tilde{\gamma}}{b\omega_2})\cos^2\theta. \nonumber \\
\label{27}
\end{eqnarray}
From (\ref{27}), we get the follwing relation between the conserved charges
\begin{equation}
E - J_1 = \frac{S}{\omega_1} + \frac{J_2^{\prime\prime}}{\omega_2} \ ,
\nonumber \\ \label{28}
\end{equation}
where $J_2^{\prime\prime} =
\frac{b\omega_2}{\sqrt{1-{\Omega_1}^2}}J_2$. For completeness we
wish to compute $J_1$ and $J_2$ as
\begin{eqnarray}
J_1 &=&
-\frac{2Ta}{a^2-b^2}~\int_{\frac{\pi}{2}}^{\theta_1}
\frac{d\theta}{\theta_y}\cos^2\theta
= \frac{2T}{\Omega_1}\cos\theta_1 \ ,
\cr & \cr
J_2 &=&
\frac{2Ta}{a^2-b^2}\sqrt{1 - \Omega_1^2}~
\int_{\frac{\pi}{2}}^{\theta_1} \frac{d\theta}
{\theta_y}\cos^2\theta
= -\frac{2T}{\Omega_1}\sqrt{1 - \Omega_1^2}\cos\theta_1 \ .
\nonumber \\ \label{29}
\end{eqnarray}
From (\ref{29}), we have the follwing relation between $J_1$ and $J_2$
\begin{equation}
J_1 = \sqrt{J_2^2 + \frac{\lambda}{\pi^2}\cos^2\theta_1} \ .
\nonumber \\ \label{30}
\end{equation}
This expression has already been found in \cite{Bobev:2007bm}. One can also
see that from (\ref{27}) we have the following relation between
$E$ and $S$,
\begin{eqnarray}
E - \frac{S}{\omega_1} = 0.
\end{eqnarray}
\subsection{Magnon solution with Four spins }

In this section, we  would like to generalize the results of the
previous section to include two spins along AdS and two angular
momenta along the deformed  S$_{\gamma}^3$. The relevant metric
and B-field is given by
\begin{eqnarray}
ds^2 &=& -\cosh^2\rho dt^2 + d\rho^2 +
\sinh^2\rho(d\psi^2 + \sin^2\psi d\xi_1^2 + \cos^2\psi d\xi_2^2)
 \cr & \cr
&&+ d\theta^2 + G\sin^2\theta d\phi_1^2 + G\cos^2\theta d\phi_2^2 \ ,
\cr & \cr
B_{\phi_1\phi_2} &=& \tilde{\gamma} G\sin^2\theta\cos^2\theta \ ,
\cr & \cr
G^{-1} &=& 1 + \tilde{\gamma}^2\sin^2\theta\cos^2\theta \ .
\nonumber \\ \label{31}
\end{eqnarray}
We take the following anstaz
\begin{eqnarray}
t &=& \tau + G_1(y), ~~ \rho = \rho(y), ~~ \psi = {\rm constant}, ~~
\xi_1 = \omega_1(\tau + G_2(y)),
\cr & \cr
\xi_2 &=& \omega_2(\tau + G_3(y)),\phi_1 =
\tau + G_4(y), ~~~ \theta = \theta(y), ~~~
\phi_2 = \omega_3(\tau + G_5(y)),
\nonumber \\ \label{32}
\end{eqnarray}
where $y = a\sigma - b\tau$.
Solving the equation of motion for the coordinates
$t, \xi_1, \xi_2, \phi_1$ and $\phi_2$,
we have the following expression for $G_1(y), G_2(y), G_3(y),
G_4(y)$ and $G_5(y)$ respectively.
\begin{eqnarray}
G_1(y) &=& \frac{1}{a^2 - b^2}\left(\frac{A_1}{\cosh^2\rho} -
b\right), \cr & \cr G_2(y) &=& \frac{1}{a^2 - b^2}
\left(\frac{A_2}{\sinh^2\rho\sin^2\psi} - b\right), \cr & \cr
G_3(y) &=& \frac{1}{a^2 - b^2}
\left(\frac{A_3}{\sinh^2\rho\cos^2\psi} - b\right), \cr & \cr
G_4(y) &=& \frac{1}{a^2 - b^2} \left(\frac{A_4}{G\sin^2\theta} - b
- a\tilde{\gamma}\omega_3\cos^2\theta\right), \cr & \cr G_5(y) &=&
\frac{1}{a^2 - b^2} \left(\frac{A_5}{G\cos^2\theta} - b -
\frac{a\tilde{\gamma}}{\omega_3}\cos^2\theta\right), \nonumber \\
\label{33}
\end{eqnarray}
where $A_1, A_2, A_3, A_4$ and $A_5$  are integration constants,
which satisfy the following relation, as derived from Virasoro
constraints,
\begin{equation}
-A_1 + \omega_1^2A_2 + \omega_2^2A_3 + A_4 + \omega_3^2A_5 = 0 \ .
\nonumber \\ \label{34}
\end{equation}
The conserved charges derived from Virasoro constraints as
explained earlier corresponding to $t, \xi_1, \xi_2, \phi_1$ and
$\phi_2$ are $E, S_1, S_2, J_1$ and $J_2$ respectively. The charges
are shown to satisfy the following dispersion relation among them
\begin{equation}
E - J_1 = \frac{S_1}{\omega_1} + \frac{S_2}{\omega_2} +
\frac{\tilde{J_2}}{\omega_3} \ , \nonumber \\ \label{35}
\end{equation}
where $\tilde{J_2} = \frac{J_2}{1 -
\frac{b\tilde{\gamma}}{a\omega_3}}$. While deriving the above relation
we have used $A_1 = b = A_4$, and $A_2 = 0 = A_3 = A_5$. One can further use
the same kind of regularization technique as discussed in previous section to get a formal
expression for the giant magnon dispersion relation in the
presence of two spins along AdS$_5$ and two angular momenta along
S$^5$. We skip the details here.

\section{Conclusions}
In this paper, we have found a class of giant magnon and single
spike string solutions in less supersymmetric real
$\beta$-deformed Lunin-Maldacena background with three spins along
various directions of AdS$_5$ and S$^5$. The relation among
conserved quantities in (\ref{16}) is similar to the undeformed
$AdS_3 \times S^3$ giant magnon relation obtained in
\cite{Ryang:2006yq}. As expected, for zero deformed parameter i.e,
$\tilde{\gamma} = 0$, we get the same value of $J_2$ and the same
relation among the charges as derived in \cite{Ryang:2006yq}.
Thus, we get the result as expected in \cite{Chu:2006ae}, where
the authors claimed that the deformed parameter should not appear
explicitly in the dispersion relation, however can be absorbed in
the definition of the conserved charges. As argued in
\cite{Chu:2006ae} if the magnon dispersion relation depends
explicitly on the deformed parameter then in general we cannot
find integrable spin chain systems. We also discarded the divergent terms in (\ref{24})
and found the regularized dispersion relation which is superposition of two magnon bound states where
the worldsheet momentum is shifted by a factor $2\pi\gamma$,
as in \cite{Chu:2006ae},\cite{Bobev:2007bm}. At this point,
it is worth mentioning about \cite{Bykov:2008bj}, where a class of magnon
solutions were derived and it was shown that in the limit of $J_2 \rightarrow
\infty$, the dispersion relation was independent of the deformation parameter.
In the present case however we would like to stress that our solutions are very similar
to the ones presented in \cite{Bobev:2007bm}, because if we switch off the spin along the
$AdS$ space we get back dispersion relation presented there. However it will be interesting to
generalize the solutions of \cite{Bykov:2008bj} to include an extra spin along the
$AdS$ direction and check the finite size correction to the magnon and spike dispersion relation.
It would also be interesting to look for the dual operators on the boundary, as one would
expect from the AdS/CFT duality. The exact nature of the operators are unknown. But the
expectations from \cite{McLoughlin:2006cg} leads us to believe that such dual operators would
exist. It would really be challenging to construct such operators dual to the spiky strings
presented in this paper.

\end{document}